\documentclass[12pt]{article}
\usepackage{mathrsfs}
\usepackage{amsmath}
\usepackage{mathrsfs}
\usepackage{amssymb}
\usepackage{color}
\newcommand{\bea}{\begin{eqnarray}}
\newcommand{\eea}{\end{eqnarray}}
\newcommand{\be}{\begin{equation}}
\newcommand{\ee}{\end{equation}}
\newcommand{\vs}[1]{\vspace{#1 mm}}

\newcommand{\bra}[1]{\langle{#1}|}
\newcommand{\ket}[1]{|{#1}\rangle}
\renewcommand{\a}{\alpha}

\renewcommand{\d}{\delta}

\newcommand{\dsl}{\pa \kern-0.5em /}
\newcommand{\la}{\lambda}
\newcommand{\half}{\frac{1}{2}}
\newcommand{\pa}{\partial}

\newcommand{\nn}{\nonumber\\}

\def\ii{{\rm i}}

\begin{document}

\topmargin 0pt \oddsidemargin 0mm

\begin{flushright}

USTC-ICTS-09-12\\

\end{flushright}

\vspace{2mm}

\begin{center}

{\Large \bf The D0$-$D8 system revisited}

\vs{10}

 {\large J. X. Lu \footnote{E-mail: jxlu@ustc.edu.cn} and Rong-Jun Wu \footnote{E-mail: rjwu@mail.ustc.edu.cn}}

\vspace{6mm}

{\em

Interdisciplinary Center for Theoretical Study\\

University of Science and Technology of China, Hefei, Anhui 230026, China\\

}

\end{center}

\vs{9}

\begin{abstract}
We address a few subtle issues regarding the interacting D0-D8
system. There are two existing interpretations for the
counter-intuitive non-vanishing  cylinder-diagram  R-R potential. We
improve them each by properly dealing with the divergence of
potential in the R-R as well as the NS-NS sector, which has been
ignored so far. We further test them by considering the D8 to carry
a flux, electric or magnetic. We find that the improved
interpretations continue to hold.  We resolve a subtle issue
regarding the regularization of fermionic zero-modes in the R-R
sector when the D8 carries an electric flux so that a meaningful
result for the potential can be calculated. The persistence of
divergence for the potential in either sector in the presence of a
flux on the D8 brane indicates that adding a flux/fluxes on the D8
brane doesn't help to improve its nature of existence as an
independent object, therefore reenforcing the previous assertion on
D8 branes.

\end{abstract}

\newpage

\section{Introduction}

Among various D$p$ branes with different dimensionality $p$, the D8
brane is peculiar\cite{polbooktwo, Polchinski:1995df,
Bergshoeff:2001pv} and difficult to understand: as a codimension one
extended object, its Ramond$-$Ramond (R-R) field does not fall off
with distance (like a planar source in 3 + 1 dimensions). As such,
when its bulk description is considered in terms of dilaton and
metric, the dilaton diverges a finite distance from the brane. It is
thus believed that the D8-brane cannot exist as an independent
object, but only in connection with orientifold planes, for example,
arising in the T-dual of  the type I theory \cite{polbooktwo}.

Part of the above nature of D8 brane also manifests itself in the
interacting D0$-$D8 system. For example, from the closed string
viewpoint, the stringy cylinder-diagram interaction energy from
 the Neveu-Schwarz$-$Neveu-Schwarz (NS-NS) sector (also from the R-R
sector) is always divergent. On the one hand, this system, like ND =
4 system\footnote{For an open string stretched between a D$p$ and a
D${p'}$(assuming $p \ge p'$), we have the string to satisfy for
every coordinate a Neumann (N) or Dirichlet (D) boundary condition
at either end. We denote by NN the number of coordinates for which
both ends have a N condition, by DD the number of coordinates for
which both ends have a D condition, and by ND  the number of
coordinates for which one end has a N condition while the other has
a D condition. Here we have the number of mixed  boundary conditions
ND = $p - p'$. A zero net static force occurs for ND = 0, 4, and 8,
respectively\cite{Polchinski:1995prl, Polchinski:1996fm,
Lifschytz:1996pl}. For the D0-D8 system, we have ND = 8 and  NN = DD
= 1. In this paper, for each given brane, we use $x^\alpha$ to label
the coordinates along the N-directions and $y^i$ to label the
coordinates along the D-directions.}, preserves  1/4 bulk
supersymmetries (susy) and as such the net static interaction
between the two branes vanishes due to the ``abstruse identity". The
zero net interaction for the ND = 4 case is, however, due to the
separately vanishing contribution from either the NS-NS sector or
the R-R sector. The vanishing NS-NS sector contribution is expected
since the two D-branes involved carry different R-R charges,
therefore giving an expected zero R-R sector contribution. However,
for the ND = 8 case at hand, we actually have an infinite NS-NS
contribution and this must imply an infinite  R-R contribution which
cancels precisely the NS-NS contribution. One special feature for
this particular system is that only the massless modes rather than
the full string spectrum appear to contribute in either sector. The
contribution from the R-R sector is however puzzling and
counter-intuitive since one naively does not expect a D0 and a D8 to
interact through a R-R gauge field\footnote{ In the open string
description, this contribution is from the R $(- 1)^F$ sector since
the two fermionic zero modes in one NN and  one DD directions cancel
the superghost zero modes but the contribution from the NS $( -
1)^F$ sector vanishes due to the  fermionic zero modes along eight
ND-directions.}.

For this, there were various efforts trying to understand the puzzle
behind\cite{Danielsson:1997wq, Bergman:1997gf, Billo:1998np}. One
way to interpret this massless contribution is to identify it as due
to a string stretched between the two branes with its tension
one-half of that of a fundamental string \cite{Danielsson:1997wq,
Bergman:1997gf}. This is supported by the existence of a coupling of
a string, with its tension one-half of that of the fundamental
string, with a worldvolume gauge field in the 8-brane effective
action in the presence of a D0 brane \cite{Danielsson:1997wq}. This
is further supported when the 0-brane in the presence of the 8-brane
is considered in the massive type IIA supergravity along the line of
\cite{Polchinski:1995sm}. This interpretation explains the anomalous
creation of a fundamental string when the 0-brane crosses the
8-brane and this string creation  is in turn related to the
Hanny-Witten effect \cite{Hanany:1996ie} by a series of dualities.
It holds also for system D$_{8 - p}$-D$_p$ which is related to the
D0-D8 system by T-duality. This approach considers so far only the
finite piece of the infinite contribution from either the NS-NS or
the R-R sector.

The other interpretation tries to follow the same footing as in ND =
0 and 4 cases. This is to identify the R-R charge carried by the
0-brane with the one opposite to the R-R charge carried by the
8-brane via a duality relation which holds only for this particular
system and those related to this by T-duality and as such the
interaction from the R-R sector is nothing but the usual attractive
Coulomb-like force between the two branes\cite{Billo:1998np}.

In this paper, we try to address how to implement these
interpretations with the consideration of the divergence of the
potential and extend them to the case when the D8 brane carries a
constant flux, electric or magnetic, along with a few subtle issues
which need clarifications. We also examine whether there is an
improvement on the D8 brane nature of existence as an independent
object when it carries a flux, electric or magnetic.

This paper is organized as follows. In section 2, we  address an
issue on how to deal with the divergence in the interaction energy
from either sector between a D0 and a D8 which has been ignored so
far in the first interpretation even though this is not a real
concern in the above second interpretation. We also calculate the
R-R Coulomb potential including the divergent piece via an effective
field theory approach where the crucial duality relation recognized
in the second interpretation is employed. In section 3, we consider
the case of D8 carrying a constant electric flux and show that the
improved versions of these interpretations continue to hold. We also
address a related subtle issue regarding the regularization of
fermionic zero-modes in the calculation of the R-R contribution. In
section 4, we consider the case of D8 carrying a magnetic flux and
show that once again the improved versions still hold even though
there is now a net force between the two branes. We also examine if
the D8 brane nature of existence as an independent object can be
improved when it carries a flux, electric or magnetic. We summarize
the results in section 5.

\section{On the divergence}
For the system under consideration and for concreteness, we assume
the D0 along $x^0$ and the D8 along $x^0, x^1, \cdots x^8$ with the
D0 located at $y^1 = y^2 = \cdots = y^8 =0, y^9 = - Y$ with $Y >
0$(see footnote 3 for notation conventions) and the D8 at $y^9 = 0$.
In other words, the two are separated by a distance $Y$ along the
9th direction. The interaction potential between the two in either
sector can be calculated either as a one-loop annulus amplitude from
the open string description \cite{Lifschytz:1996pl} or as a
tree-level cylinder amplitude from the closed string boundary state
description \cite{Billo:1998np}. The contribution of the NS and R
open string sectors to the potential per unit D0 brane worldvolume
in the NS-NS sector is \be\label{nsnsp} V_{\rm NS-NS} (Y) =
\frac{1}{2} \,(8 \pi^2 \alpha')^{-1/2} \int_0^\infty dt \, e^{-
\frac{Y^2 t}{2\pi\alpha'}} \, t^{- 3/2}, \ee which gives rise to a
finite constant repulsive force acting on the D0 as \bea
\label{nsnsforce} F_{\rm NS-NS} = \frac{d V_{\rm NS-NS} }{d
Y\,\,\,\,\,\,\,\,} = - \frac{1}{(4\pi)^{1/2}} \frac{1}{2\pi\alpha'}
\int_0^\infty d x\, e^{- x}\, x^{- 1/2} = - \frac{1}{4 \pi\alpha'} =
- \frac{T_0}{2}, \eea where we have changed the integration variable
$t$ to $x = Y^2 t /2\pi\alpha'$ in the second equality and the
resulting integration is simply the well-defined gamma-function
$\Gamma (1/2)$, and $\alpha'$ is the string constant, related to the
tension of a fundamental string via $T_0 = (2\pi \alpha')^{-1}$. The
contribution of R $(- 1)^F$ open string sector to the cylinder
potential per unit D0-brane worldvolume in the R-R sector is
\be\label{rrp} V_{\rm R-R} (Y) = - \frac{1}{2} \,(8 \pi^2
\alpha')^{-1/2} \int_0^\infty dt \, e^{- \frac{Y^2 t}{2\pi\alpha'}}
\, t^{- 3/2}, \ee which gives rise to an finite attractive constant
force acting on the D0 brane. Similarly, this  force can be obtained
as \bea \label{rrforce} F_{\rm R-R} = \frac{d V_{\rm R-R} }{d
Y\,\,\,\,\,\,\,\,} = \frac{T_0}{2}.\eea The peculiar feature for
this particular system is that only the massless modes rather than
the full string spectrum appear to contribute to the potential in
either sector. In addition, the force from either sector is constant
as shown above. As expected, due to the preservation of 1/4 susy,
the total potential or the net force acting on either object in the
system vanishes as obviously from the above. The non-vanishing R-R
contribution can either be inferred from the `no-force' condition
with the easily calculated NS-NS contribution for this system or can
be calculated, using a regularization of fermionic zero-modes,  from
the one-loop open string annulus diagram in the $R (-1)^F$ sector
\cite{Lifschytz:1996pl} or from the tree-level closed string
cylinder diagram in the R-R sector \cite{Billo:1998np}. In spite of
this, it is puzzling and difficult to understand this result as
discussed in the Introduction.

To resolve this puzzle, one way is to interpret the constant
attractive R-R force given in Eq.(\ref{rrforce}) above as due to a
string stretched between the D0 and the D8 with its tension one half
of that of a fundamental string \cite{Danielsson:1997wq,
Bergman:1997gf}. This interpretation is in line with
\cite{Polchinski:1995sm} for a D0 brane in the presence of a D8
brane and is also consistent with the Hanny-Witten effect
\cite{Hanany:1996ie} for a D0 crossing a D8. However, there is a
subtle issue regarding the divergence of potential in either sector
given above which has been ignored so far, for example, in
\cite{Bergman:1997gf} when such an interpretation is invoked. The
potential in either sector as given above contains the following
integration \be \label{int} \int_0^\infty dt \, e^{- \frac{Y^2
t}{2\pi\alpha'}} \, t^{- 3/2} = \frac{Y}{(2\pi\alpha')^{1/2}}
\int_0^\infty d x \, e^{- x} \, x^{- 3/2},\ee which is actually
divergent and where on the right we have used the integration
variable $x = Y^2 t/2\pi\alpha'$. Note that the integration on the
right above is not simply the gamma function $\Gamma (- 1/2)$ as
usually taken in the literature, for example, in
\cite{Bergman:1997gf}. The  integral representation of $\Gamma (z)$
with $z$ a complex number \be \Gamma (z) = \int_0^\infty dx \, e^{ -
x} x^{z - 1},\ee is valid only for ${\rm Re}\,(z) > 0$. When ${\rm
Re} \,(z) \le 0$, $\Gamma (z)$ can still be defined but it doesn't
have the above integral representation. One has to use its other
representations. In other words, the integration on the right of
Eq.(\ref{int}) is actually divergent, not the finite $\Gamma (-
1/2)$. Due to the divergent nature, we need to deal with the
integration carefully. For this let us begin with the left side of
(\ref{int}) and denote the integration as $I$ \bea I \equiv
\int_0^\infty dt \, e^{- \frac{Y^2 t}{2\pi\alpha'}} \, t^{- 3/2} = -
2 \left. t^{-1/2}\, e^{- \frac{Y^2 t}{2\pi\alpha'}} \right|_0^\infty
- \frac{2 Y}{(2\pi\alpha')^{1/2}} \int_0^\infty d x \, e^{- x} \,
x^{- 1/2}, \eea where we have performed an integration by part and
changed the integration variable to $x = Y^2 t/2\pi\alpha'$ in the
second term. The first term is $\infty$, denoted as $I_\infty$. The
integration in the second term is nothing but the well-defined
$\Gamma (1/2) = \sqrt{\pi}$. Note that the divergence of $I_\infty$
is due to $t = 0$ at which the exponential in the first term becomes
unity and we therefore expect $I_\infty$ to be independent of $Y$.
This is consistent with the fact that the force calculated in either
sector above is constant and can indeed be obtained merely from the
second term. So we have now $I = I_\infty - \frac{2 \pi^{1/2}
Y}{(2\pi\alpha')^{1/2}}$, i.e., a separation-independent divergent
term plus a separation-dependent finite piece. With this, we have
\bea \label{nsns-rr-f} V_{\rm NS-NS} (Y) = \frac{I_\infty}{2 (8\pi^2
\alpha')^{ 1/2}} - \frac{Y}{4\pi \alpha'}, \qquad V_{\rm R-R} (Y) =
- \frac{I_\infty}{2 (8\pi^2 \alpha')^{ 1/2}} + \frac{Y}{4\pi
\alpha'}.\eea So in addition to the finite piece, the potential in
each sector has a divergent piece which is independent of the brane
separation. It is obvious that this divergent piece is the
contribution between the D0 and the D8 when the two is put on top of
each other, i.e., with a zero separation. So this piece in either
sector can be viewed as the corresponding zero-point energy. When
the two move away from each other, there is an additional finite
piece created which can be viewed as the Casimir energy of the
system in either sector. In the R-R sector, this can be taken as due
to the creation of a string stretched between the two with its
tension one half of that of the fundamental string. Note that the
divergence must come from the nature of D8 brane as mentioned in the
Introduction. So whenever this persists, the corresponding system
cannot be taken as an independent object.

With this divergent piece in either  sector, how then can we make a
consistent picture using the creation of a fundamental string when
the D0 crosses  the D8 adiabatically? For this, we need to use the
improved version of this interpretation given in \cite{Kitao:1998vn}
where it is stressed that such a crossing is merely a parity
transformation along the DD direction (i.e., along the $x^9$
direction). Such a transformation will reverse the sign of $R ( -
)^F$ term but leave the rest of terms invariant, therefore not
converting a brane into an anti-brane in general even though it
doesn't make difference for the present case since we have the $NS
(-)^F$ term vanishing. If the initial configuration is
supersymmetric as the present case and is taken as the vacuum
configuration, we have the total potential as $ NS + R + NS (-)^F +
R (-)^F$, where the first two terms give the NS-NS sector potential
while the last two terms give the R-R sector one. After the
adiabatically crossing, we end with a vacuum configuration $NS + R +
NS (-)^F - R(-)^F$ since only the $R (-)^F$ will change its sign,
which is obviously non-supersymmetric unless the $R (-)^F$ term
vanishes. But this is an adiabatic process and the total energy
should be kept unchange, therefore we must have created an
additional term $2 R (-)^F$ in the process, as argued in
\cite{Kitao:1998vn}, such that \be NS + R + NS (-)^F + R (-)^F =
\left( NS + R + NS (-)^F - R (-)^F\right) + 2 R (-)^F.\ee If the
divergent piece in either sector is not considered, then the above
is consistent with the creation of a fundamental string since the
term $2 R(-)^F$ is just the tension of a fundamental string times
the separation. For the present case, the above continues to hold if
we include the divergent piece in either sector and the
interpretation is now that the crossing creates not only a
fundamental string but also twice the $R (-)^F$ sector zero-point
energy before crossing whose origin may be better explained using
the second interpretation proposed in \cite{Billo:1998np} when the
brane separation vanishes.  So we further improve the improved
interpretation proposed in \cite{Kitao:1998vn} here.

Note that we don't need to improve the second interpretation given
in \cite{Billo:1998np} much except for merely insisting that the
Coulomb-like  potential includes the divergent piece as well as the
finite piece (so does the potential in the NS-NS sector) for which
we turn next. This R-R interaction is purely due to massless modes
and as mentioned earlier, just like the field theory limit of the ND
= 0 case, it is simply due to the usual Coulomb-like force between
the two D-branes. This is however counter-intuitive for the ND = 8
system and any other system related to this by T-duality. The
complete discussion given \cite{Billo:1998np} in terms of relevant
boundary states and vertex operators is lengthy and here we will use
the effective field theory approach, for example, following
\cite{Lu:2009yx}, to derive this same interaction based on the
underlying physics and the duality relation given there. The key for
this is the recognition that the usually completely decoupled
unphysical degrees of freedom such as the longitudinal and scalar
states, the analogues of the scalar and longitudinal photons of
electrodynamics, in perturbative string theory becomes important
when dealing with the non-perturbative boundary states. This
consideration will give rise to the duality relation, mentioned
earlier, which is crucial for the interaction to arise.

Since only the massless modes in type IIA are relevant here, the
interaction in the R-R sector is due to the exchange of an off-shell
closed string between the two R-R charges, just like the usual
Coulomb interaction between two static charges as due to the
exchange of a virtual photon. As discussed in \cite{Billo:1998np},
this off-shell closed string state can be represented by the
corresponding vertex operator which can be constructed from the
on-shell massless closed string vertex operator and by allowing the
momentum $k^2 \neq 0$, i.e., off-shell. As described in
\cite{Billo:1998np}, the R-R sector boundary state used in the
cylinder-diagram calculation is in the $(-1/2, -3/2)$ picture and to
soak up the superghost number anomaly, it can only couple to states
that are also in the asymmetric picture. However, the vertex
operators used usually for constructing the R-R states are in the
symmetric $(-1/2, -1/2)$ picture and are given as \be
\label{svertex} V_R (k; z, {\bar z}) = \frac{(C \Gamma^{\mu_1 \mu_2
\cdots \mu_{m + 1}})_{\alpha\dot\beta}}{2 \sqrt{2} (m + 1)!}
F_{\mu_1\mu_2 \cdots \mu_{m + 1}} V^\alpha_{- 1/2} (k/2; z) {\tilde
V}^{\dot\beta}_{-1/2} (k/2; \bar z),\ee with m odd (even) in IIA
(IIB) theory and $V^\alpha_l (k; z) = c (z) S^\alpha (z) e^{l
\phi(z)} e^{i k\cdot X (z)}$. The $(m + 1)$-form $F_{m + 1}$ gives
just the R-R field strength of IIA or IIB theory since the BRST
invariance of the vertex operator requires $k^2 = 0$, and $d F_{m +
1} = d \ast F_{m + 1} = 0$ which are precisely the Bianchi identity
and the field equations of motion for the field strength.

The vertex operators in the $(-1/2, -3/2)$ can also be constructed
and they can be transformed to the ones in the symmetric picture via
a picture changing operation in the right sector
\cite{Friedan:1985ge}. For the present purpose, we don't need the
complete construction as given in \cite{Billo:1998np} but the
following term \cite{Bianchi:1991eu} \be \label{asvertex} W^{(0)}
(k; z, \bar z) = \frac{(C \Gamma^{\mu_1 \cdots
\mu_m})_{\alpha\dot\beta}}{m !} A_{\mu_1\cdots\mu_m} V^\alpha_{-1/2}
(k/2; z) {\tilde V}^{\dot\beta}_{- 3/2} (k/2; \bar z).\ee Note that
instead of the R-R field strength as in the symmetric picture, we
need here only the R-R m-form gauge potential $A_m$. The BRST
invariance of this vertex requires $k^2 = 0$ and $d A_m = d \ast A_m
= 0$, therefore just a pure gauge potential. But this is sufficient
for what follows.

    In the symmetric case, the electric-magnetic duality is $F_{m +
    1} \simeq \ast F_{9 - m}$. If we consider the spatial momentum in
    (\ref{svertex}) only along the 9th direction, then we have from
    the duality relation \be F_{01} = k_0 A_1 = - F_{23\cdots 9} =
    k_9 A_{2\cdots 8}.\ee Since the state considered is massless,
    therefore $k_0 = k_9$ and this gives \be A_1 = A_{2\cdots 8}.\ee
    It is important to realize that the above relation involves only
    the physical degrees of freedom.

    We now move to the asymmetric state (\ref{asvertex}). As will be
    seen, the above given conditions for keeping the state BRST invariant
    are precisely those fulfilled by the mixture of longitudinal
    and scalar polarizations which describe the Coulomb interaction. Let
    us focus first on the 1-form potential and consider again the
    spatial momentum along 9th direction. Then the pure gauge
    solution is simply, with $k_0 = k_9$, as \be \label{oneform} A_0 = A_9, \quad
    {\rm and} \quad A_i = 0, \quad {\rm for}\quad i = 1, 2, \cdots 8.\ee
    For the state (\ref{asvertex}), we have a Hodge duality for the
    10-dimensional unphysical polarizations. For 1-form potential,
    we have $A_9 = - A_{01\cdots 8}$. When combined this with
    (\ref{oneform}), we have \be \label{duality} A_0 = - A_{01\cdots 8}.\ee
    This unusual relation is of no relevance in perturbative string
    theory where the unphysical degrees of freedom always decouple
    but it has remarkable consequences for the present case. In fact,
    it implies that the charge felt by $A_0$ is opposite to that
    felt by $A_{01\cdots 8}$ and thus the attractive Coulomb R-R
    force between a D0 and a D8 brane can be understood as due to
    the exchange of longitudinal and scalar polarizations identified
    according to (\ref{duality}). In the following, we will use this
    relation to derive the R-R Coulomb interaction via the
    field theory approach \cite{Lu:2009yx}.

    With the canonical normalization \cite{Lu:2009yx} for the background R-R potential,
    the coupling for a D0 with the 1-form R-R potential $A_0$ is
    \be J^{(0)} = \sqrt{2} \,c_0\, V_1\, A_0, \ee
    and the coupling for a D8 with the 9-form R-R potential is
    \be J^{(8)} = \sqrt{2}\, c_8\, V_9\, A_{01\cdots 8},\ee
    where $V_{p + 1}$ is the worldvolume of $D_p$
    brane and the constant $c_p = \sqrt{\pi} (2\pi \sqrt{\alpha'})^{3 - p}$.
    Then the static interaction per unit D0 brane worldvolume due to the massless modes in the
    R-R sector can be calculated in the momentum space simply as
    \be \label{rrpm} V_{\rm R-R} (k_\bot) = - \frac{1}{V_1 V_9} \underbrace{J^{(0)}
    J^{(8)}} = - 2 \, c_0\, c_8 \underbrace{A_0\, A_{01\cdots 8}} = - \frac{1}{2\pi \alpha' \, k^2_\perp}, \ee
    where we have used the explicit expression of $c_p$ and the key fact \be \underbrace{A_0 \, A_{01\cdots 8}} = -
    \underbrace{A_0\, A_0} = - \underbrace{A_{01\cdots 8}\,A_{01\cdots
    8}} = \frac{1}{k^2_\perp} \neq 0,\ee because of (\ref{duality}). In the above,
     the propagator $\underbrace{A_{0\cdots p} \, A_{0\cdots p}} = -
    1/k^2_\perp$ is used and the
    $k_\perp$ is the spatial momentum perpendicular to both
    branes, i.e., the momentum along the 9-th direction for the present
    case. So the potential in coordinate space is \bea V_{\rm R-R}
    (Y)
    &=&
    \int_{-\infty}^\infty \frac{d k_\perp}{2\pi} e^{- i k_\perp Y}
    V_{\rm R-R} (k_\perp)\nn
    &=& - \frac{2}{(2\pi)^2 \alpha'} \int_0^\infty d k_\perp
    \frac{\cos k_\perp Y}{k^2_\perp}\nn
    &=& \frac{2}{(2\pi)^2 \alpha'} \left( \left.\frac{\cos k_\perp
    Y}{k_\perp}\right|_0^\infty  + Y \int_0^\infty \frac{\sin
    k_\perp Y}{k_\perp} d k_\perp \right)\nn
    &=& - \frac{I_\infty}{2 (8\pi^2 \alpha')^{ 1/2}} + \frac{Y}{4\pi
\alpha'},
    \eea
  where in obtaining the last line we have defined $k_\perp = \sqrt{2 t/\pi^2 \alpha'}$ in the first term
  and made use of $\int_0^\infty d x \sin x /x = \pi/2$ in the second term in the bracket in the third line above.
  This potential is identical to the stringy result given earlier in
  (\ref{nsns-rr-f}) but is now calculated via the field theory approach
  as the Coulomb interaction. As stressed, the infinite piece should
  be kept which reflects the nature of the D8 brane and it is just the
  Coulomb interacting energy when the brane separation vanishes.

\section{The D8 with an electric flux}

In this section, we will examine whether the improved versions of
the two interpretations for the non-vanishing R-R potential
discussed in the previous section in the absence of a flux continue
to hold when the D8 carries an electric flux. We will also discuss a
subtle issue regarding the regularization of fermionic zero-modes in
the calculation of the R-R potential for the present case.

For this, we first calculate the potential in the NS-NS sector. This
can be worked out in almost the same way as the case when the D8
doesn't carry any flux \cite{Lifschytz:1996pl, Billo:1998np} if a
trick as adopted in \cite{Lu:2009yx,Lu:2009se, Lu:2009th} is
employed. In the following, we will perform the calculations using
the closed string operator formalism in which D-branes with/without
constant fluxes can be described by the so-called boundary states.
In this approach, there are two sectors, namely NS-NS and R-R
sectors, respectively. For later convenience, we give a brief
description of each and refer to
\cite{Billo:1998np,Lu:2009yx,Lu:2009se} for detail.

Both in the NS-NS and R-R sectors, there are two possible
implementations for the boundary conditions of a D-brane which
correspond to two boundary states $|B, \eta \rangle$ with $\eta =
\pm $. However, only the so-called Gliozzi-Scherk-Olive (GSO)
combinations $\ket{B}_{\rm NS-NS} = \frac{1}{2} \left[
\ket{B,+}_{\rm NS-NS} - \ket{B,-}_{\rm NS-NS} \right]$ and
$\ket{B}_{\rm R-R} = \frac{1}{2} \left[ \ket{B,+}_{\rm R-R} +
\ket{B,-}_{\rm R-R} \right]$ are selected in the NS-NS and in the
R-R sectors, respectively. The boundary state $\ket{B,\eta}$ is the
product of a matter part and a ghost part as $\ket{B, \eta } =
\frac{c_p}{2}\ket{B_{\rm mat}, \eta } \ket{B_{\rm g}, \eta}$ with
$\ket{B_{\rm mat}, \eta} = \ket{B_X} \ket{B_{\psi}, \eta}$,
$\ket{B_{\rm g}, \eta} = \ket{B_{\rm gh}} \ket{B_{\rm sgh}, \eta}$
and the overall normalization $c_p = \sqrt{\pi}
(2\pi\sqrt{\alpha'})^{3 - p}$. The boundary states for ghosts and
superghosts are independent of the fluxes and therefore remain the
same as before. We will not list them here for simplicity. The
expressions of the matter part of $\ket{B,\eta}$ are given,
respectively, as $\ket{B_X} = \exp\left[-\sum_{n=1}^\infty
\frac{1}{n}\a_{-n}\cdot S\cdot \tilde \a_{-n}
\right]\ket{B_X}^{(0)}$, and $\ket{B_\psi,\eta}_{\rm NS-NS} = -\ii\,
\exp\left[\ii\,\eta\sum_{m=1/2}^\infty \psi_{-m}\cdot S\cdot\tilde
\psi_{-m}\right] \ket{0}$ for the NS-NS sector, and
$\ket{B_\psi,\eta}_{\rm R-R}$ $ = -
\exp\left[\ii\,\eta\sum_{m=1}^\infty \psi_{-m} \cdot S\cdot\tilde
\psi_{-m}\right] \ket{B,\eta}_{\rm R-R}^{(0)}$ for the R-R sector.
They look the same in form as their correspondences in the absence
of fluxes. All the information about the constant flux $F$ is
encoded in the S-matrix \be S = \left(\left[(\eta - \hat{F})(\eta +
\hat{F})^{-1}\right]_{\alpha\beta},  - \delta_{ij}\right), \ee and
the zero-mode boundary states: for the bosonic sector as\be
\label{zeromodex} |B_X\rangle^{(0)} = \sqrt{- \det \left(\eta + \hat
F\right)} \,\delta^{9 - p} (q^i - y^i) \prod_{\mu = 0}^9 |k^\mu =
0\rangle,\ee  and for the R sector as  \be \label{zeromodep}
|B_\psi, \eta\rangle_{\rm R}^{(0)} = \left(C \Gamma^0 \Gamma^1
\cdots \Gamma^p \frac{1 + {\rm i}\, \eta \Gamma_{11}}{1 + {\rm i}\,
\eta } U \right)_{AB} |A\rangle |\tilde B\rangle. \ee  In the above,
we have denoted by $y^i$ the positions of the D-brane along the
transverse directions, by $C$ the charge conjugation matrix and by
$U$ the following matrix \be \label{matrixu} U = \frac{1}{\sqrt{-
\det (\eta + \hat F)}} ; {\rm exp}\left(- \frac{1}{2} {\hat
F}_{\alpha\beta}\Gamma^\alpha\Gamma^\beta\right);\ee where the
symbol $;\quad ;$ means that one has to expand the exponential and
then to anti-symmetrize the indices of the $\Gamma$-matrices.
$|A\rangle |\tilde B\rangle$ stands for the spinor vacuum of the R-R
sector. We also define $\hat F = 2\pi \alpha' F$ in the above. We
would like to point out that the $\eta$ in the above means either
sign $\pm$ or the flat signature matrix $(-1, +1, \cdots, +1)$ on
the world-volume and should not be confused from the content.

Note that D0 cannot carry any flux and  without loss of generality,
we can always choose the constant electric flux  on the D8 with the
only non-vanishing components $(\hat F)_{01} = - (\hat F)_{10} = -
f$.  We then have the S-matrix for D0 and D8, respectively, as \be
\label{smatrix} S^0_{\mu\nu} = - \delta_{\mu\nu},\,\,\,\,
S^8_{\mu\nu} = \left(S^8_{\alpha\beta} \,,\, - 1\right), \ee with
the only non-vanishing components of $S^8_{\alpha\beta}$ as
$S^8_{00} = - S^8_{11} = - \frac{1+f^2}{1-f^2}, S^8_{01} = -
S^8_{10} = -\frac{2f}{1-f^2}$, and $S^8_{aa} = 1$ where $a = 2,
3\cdots 8$. With the above preparation and by considering the
relevant conventions described in footnote 3, the tree-level
cylinder-diagram amplitude in the NS-NS sector can be carried out
straightforwardly and is given as \bea\label{nsnsampe} \Gamma_{\rm
NS-NS} &=& _{\rm NS-NS}\bra{B^0}D\ket{B^8}_{\rm NS-NS},\nn &=&
\frac{\sqrt{1-f^2}}{2}\,n_0\,n_8\,V_1\,(8\pi^2\a')^{-\frac{1}{2}}
\int_0^\infty
dt\,e^{-\frac{Y^2}{2\pi\a't}}\,t^{-\frac{1}{2}}\left[\frac{\theta_4^4\left(0|\,\ii\,t\right)}{\theta_2^4\left(0|\,\ii\,t\right)}
-
\frac{\theta_3^4\left(0|\,\ii\,t\right)}{\theta_2^4\left(0|\,\ii\,t\right)}\right],\nn
&=&
-\frac{\sqrt{1-f^2}}{2}\,n_0\,n_8\,V_1\,(8\pi^2\a')^{-\frac{1}{2}}
\int_0^\infty dt\,e^{-\frac{Y^2}{2\pi\a't}}\,t^{-\frac{1}{2}},
 \eea
where $D$ is the standard closed string propagator and in the last
step we have used the usual Jacobi's ``abstruse identity'' $
\theta_3^4\left(0|\,\ii\,t\right)-\theta_4^4\left(0|\,\ii\,t\right)-\theta_2^4\left(0|\,\ii\,t\right)=0$.
In the above, we have replaced the normalization constant $c_k$ ($k
=0, 8$) by $n_k c_k$ with $n_k$ an integer to count the multiplicity
of the corresponding branes. In carrying out the above calculations,
as mentioned at the beginning of this section, we follow the trick
as adopted in \cite{Lu:2009yx,Lu:2009se, Lu:2009th} by making a
respective unitary transformation of the oscillators in the boundary
state for the D0 such that its $S^0$-matrix completely disappears
while the boundary state for the D8 ends up with a new $S = S^8
(S^0)^T$ where $T$ denotes the transpose. This new S-matrix shares
the same property as the original $S^k$ satisfying
$((S^k)^T)_\mu\,^\rho (S^k)_\rho\,^\nu = \delta_\mu\,^\nu$ with $k =
0\, {\rm or}\, 8$ but its determinant is always unity and therefore
can always be diagonalized to give its eigenvalues. With this trick,
the amplitude calculations as mentioned earlier are no more
complicated than the case for the D8 carrying no flux.

The above amplitude gives a repulsive potential per unit D0 brane
worldvolume as \be V_{\rm NS-NS} =
\frac{\sqrt{1-f^2}}{2}\,n_0\,n_8\,\,(8\pi^2\a')^{-\frac{1}{2}}
\int_0^\infty dt\,e^{-\frac{Y^2}{2\pi\a't}}\,t^{-\frac{1}{2}}, \ee
which differs from the one for the D8 carrying no flux only by an
overall factor $\sqrt{1-f^2}$. Apart from this, the electric flux on
the D8 doesn't change the structure of the potential at all. This is
simply due to the fact that the eigenvalues of the above mentioned
S-matrix, which determine the amplitude structure, remain the same
as if there is no such an electric flux present. In fact, the
eigenvalues of the matrix $S_\mu\,^\nu$ have two of $ + 1$ and eight
of $- 1$ as one can check easily and explicitly. This overall factor
arises from the overall factor appearing in the bosonic zero-mode
boundary state given in (\ref{zeromodex}).

The underlying physics of the above can be understood in the
following: If we T-dualize  along the electric flux direction on the
D8, then the D8 will become a D7 moving with a velocity of magnitude
$|f|$ ( $ \le 1$) in the original flux direction while the D0
becomes a D1 along the original flux direction but at rest. We can
make a boost $\gamma = 1/\sqrt{1 - f^2}$ against the D1 direction
such that the D7 becomes at rest. Note that this boost has no
influence on the D1 since it has a Lorentz symmetry on its
worldvolume. So after the boost, both the D1 and the D7 are at rest.
But such a boost has an effect on the tension and the R-R charge of
the D7, reducing each by a factor of $\sqrt{1 - f^2}$, since such a
boost is in the direction orthogonal to its worldvolume. Now the D1
and the D7 are orthogonal to each other and are related to the D$_{8
- p}$-D$_p$ system by T-duality, which preserves 1/4 of
supersymmetries. In particular, we can T-dualize the above D1-D7
back to a static D0-D8 system with the D8 carrying no flux but with
its tension and R-R charge $\sqrt{1 - f^2}$ times those of a
fundamental D8, respectively. Such an understanding gives not only a
physical explanation to the above calculated amplitude but also
predicts that the D0-D8 system preserves the same 1/4 of susy
whether the D8 carries an electric flux or not. This latter must
imply, due to the `no-force' condition, the amplitude in the R-R
sector, \be \label{rrap} \Gamma_{\rm R-R} =
\frac{\sqrt{1-f^2}}{2}\,n_0\,n_8\,V_1\,(8\pi^2\a')^{-\frac{1}{2}}
\int_0^\infty dt\,e^{-\frac{Y^2}{2\pi\a't}}\,t^{-\frac{1}{2}}, \ee
and the corresponding attractive potential per unit D0 brane
worldvolume  \be V_{\rm R-R} (Y) = -
\frac{\sqrt{1-f^2}}{2}\,n_0\,n_8\,\,(8\pi^2\a')^{-\frac{1}{2}}
\int_0^\infty dt\,e^{-\frac{Y^2}{2\pi\a't}}\,t^{-\frac{1}{2}}.\ee
Once again, only the massless modes rather than the full spectrum
contribute to the interaction in either sector.

With the above known answer, we now try to calculate the amplitude
in the R-R sector directly and correctly. The complication arises
from the regularization of fermionic zero-modes which requires care.
The tree-level cylinder-diagram amplitude in this sector can be
similarly calculated up to
 \bea\label{AmpRE}
\Gamma_{\rm R-R} &=& _{\rm R-R}\bra{B^0}D\ket{B^8}_{\rm R-R} \nn &=&
\frac{\sqrt{1-f^2}}{2^5}n_0\,n_8\,V_1\,(8\pi^2\a')^{-\frac{1}{2}}
\int_0^\infty dt\,e^{-\frac{Y^2}{2\pi\a't}}\,t^{-\frac{1}{2}} ~
^{(0)}_{\rm R-R}\langle B^0,\eta^0 | B^8,\eta^8\rangle_{\rm
R-R}^{(0)},
 \eea
where ${_{\rm R-R}^{\,\,\,\,\,\,(0)}\langle} B^0,\eta^0 |
B^8,\eta^8\rangle_{\rm R-R}^{(0)}= {_{\rm
R-R}^{\,\,\,\,\,\,(0)}\langle} B_\psi^0,\eta^0 |
B_\psi^8,\eta^8\rangle_{\rm R-R}^{(0)} ~ {_{\rm
R-R}^{\,\,\,\,\,\,(0)}\langle} B_{\rm sgh}^0,\eta^0 | B_{\rm
sgh}^8,\eta^8\rangle_{\rm R-R}^{(0)}$. In extracting a finite
meaningful result from the two divergent matrix elements due to the
fermionic zero-modes and the superghosts, respectively,  we need to
regularize both properly. The regularization adopted in
\cite{Bergman:1997gf, Billo:1998np} is however not directly
applicable at present since the matrix $U$ (\ref{matrixu}) appearing
in the R-R zero-mode boundary state (\ref{zeromodep}) in the
presence of an electric flux mixes a NN-direction $\Gamma^0$-matrix
with a ND-direction $\Gamma^1$-matrix. As a result, it is not
possible to group pairs of $\Gamma$-matrices purely from the
ND-directions and purely from NN or DD directions. In other words,
if we try to group, there is at least one pair with one
$\Gamma$-matrix from a ND-direction and the other from a NN- or
DD-direction. Such a grouping cannot give a meaningful result as
stressed in \cite{Billo:1998np}. One simple reason is that this
grouping is not consistent with T-duality since the corresponding
SO(2) rotation can convert a ND-direction to a NN or DD-direction
and vice-versa but T-duality doesn't allow a ND-direction to become
a NN- or DD-direction and vice-versa.

So we face a dilemma here.  The evaluation of
$^{\,\,\,\,\,\,(0)}_{\rm R-R}\langle B^0,\eta^0 |
B^8,\eta^8\rangle_{\rm R-R}^{(0)}$ ends up with two terms ${\rm
Tr}(\Gamma^0 \Gamma^9) - f {\rm Tr}(\Gamma^1 \Gamma^9)$ with ${\rm
Tr}$ denoting the trace, where the second term associated with the
flux doesn't allow a meaningful regularization since `1' is a
ND-direction while `9' is a DD-direction. Our previous discussion on
how to get rid of the electric flux on the D8 via a T-duality and a
boost motivates us to resolve this, however. These transformations
shouldn't change the value of matrix element as we discussed above
for the evaluation of the amplitude in NS-NS sector. So we can
evaluate this matrix element after the above transformations. A
T-duality doesn't change the direction of `1' but a boost will
rotate `0' and `1'. This motivates us to define new ${\tilde
\Gamma}^0$ and ${\tilde \Gamma}^1$ in terms of the old $\Gamma^0$
and $\Gamma^1$ as
 \bea\label{gammamatrixt}
\tilde \Gamma^0 =
\frac{1}{\sqrt{1-f^2}}\left(\Gamma^0-f\Gamma^1\right) ~~{\rm
and}~~\tilde \Gamma^1 =
\frac{1}{\sqrt{1-f^2}}\left(-f\Gamma^0+\Gamma^1\right),
 \eea
where the new $\tilde \Gamma^0$ and $\tilde \Gamma^1$ satisfy their
respective own properties and further $\tilde \Gamma^0\tilde
\Gamma^1 = \Gamma^0 \Gamma^1$. Using the newly  defined
$\Gamma$-matrices, we have $ ^{\,\,\,\,\,\,(0)}_{\rm R-R}\langle
B_\psi^0,\eta^0 | B_\psi^8,\eta^8\rangle_{\rm R-R}^{(0)} =
-\d_{\eta^0\eta^8,-}{\rm Tr}\left(\tilde
\Gamma^{0}\Gamma^{9}\right)$, where the trace can now be regularized
following the regularization procedure given in \cite{Billo:1998np}
since the $\tilde \Gamma^0$ is one along a NN-direction and
$\Gamma^9$ is one along a DD-direction. It is then straightforward
to have $ ^{\,\,\,\,\,\,(0)}_{\rm R-R}\langle B^0,\eta^0 |
B^8,\eta^8\rangle_{\rm R-R}^{(0)} = 2^4$.

So we have from (\ref{AmpRE})
 \bea\label{AmpRE001}
\Gamma_{\rm R-R} =
\frac{\sqrt{1-f^2}}{2}n_0\,n_8\,V_1\,(8\pi^2\a')^{-\frac{1}{2}}
\int_0^\infty dt\,e^{-\frac{Y^2}{2\pi\a't}}\,t^{-\frac{1}{2}},
 \eea
which is identical to the predicted one given in (\ref{rrap}). This
in turn shows that our above well-motivated regularization process
makes sense.

The underlying physical picture given earlier in this section also
makes it certain that the two improved interpretations discussed in
the previous section continue to apply to the present system. This
has to be true if these interpretations make sense indeed since the
present system preserves the same number of susy as the system
discussed in the previous section for which the D8 doesn't carry any
flux. We know that the present system is equivalent to a D0-D8
system with the D8 carrying no flux but with its tension and R-R
charge reduced by a factor of $\sqrt{1 - f^2}$ but the corresponding
quantities for the D0 remain untouched. This clearly indicates that
the improved second interpretation regarding the R-R interaction as
purely due to the usual attractive Coulomb-like force between the
two branes\cite{Billo:1998np} continues to hold. Also with the above
in mind, it is obvious that the improved first interpretation holds
also if one notices that the tension of the string stretched should
also be reduced by this same factor since the string is along a
DD-direction and its tension is energy per unit length and so it
should be re-scaled by this same factor under the boost. This can
also be equivalently inferred from the scaling of the D8-brane
tension from either the R-R potential or the NS-NS potential plus
the `no-force' condition since it is just the finite piece described
in the previous section.

\section{The D8 with a magnetic flux}

The D8 carrying a constant electric flux as discussed in the
previous section is actually the so-called (F, D8) non-threshold
bound state. The zero net force between a D0 and a (F, D8) can also
be intuitively understood by noticing that the D0 doesn't interact
with either constituent in the bound state. When a D8 carries a
constant magnetic flux, it actually represents the so-called (D6,
D8) non-threshold bound state. But now since the D0 has a net
repulsive interaction with the D6 in the bound state
\cite{polbooktwo}, we expect that there is a repulsive non-zero net
force between the D0 and the (D6, D8). So this system cannot
preserve any susy in general\footnote{In certain special cases as
specified in \cite{Witten:2000mf}, this system can be supersymmetric
and we will not discuss these special cases here. See also
\cite{Mihailescu:2000dn, Fujii:2001wp}} and one also expects that
the interaction is no longer purely due to the massless modes as
would be the case for pure D8 or for D8 carrying an electric flux.
However, this latter point applies only to the NS-NS sector
interaction  since there is no R-R sector interaction between the D0
and the D6. We therefore expect that the R-R sector amplitude
remains the same as in the absence of the magnetic flux which will
be shown shortly. This implies that our improved interpretations
given in section 2 continue to hold even in this case. Let us now
calculate this non-vanishing interaction which can be performed even
simpler than the previous case since we don't have a regularization
issue here for the fermionic zero modes in the R-R sector.

We also have the S-matrix for D0 or D8 to be given in
(\ref{smatrix}) but with now the only non-vanishing  components of
$S^8_{\alpha\beta}$ as $ - S^8_{00} = S^8_{11} = \cdots = S^8_{66} =
1, S^8_{77} = S^8_{88} = \frac{1-f^2}{1+f^2}, S^8_{78} = - S^8_{87}
= \frac{2f}{1+f^2}$. For this, we have taken the only non-vanishing
components of the magnetic flux as $\hat F_{78} = - \hat F_{87} = -
f$. With the same trick as mentioned in the previous section, the
new matrix $S = S^8 (S^0)^T$ has its eigenvalues as
$\{\la,\,\la^{-1},\, 1,\,1,\,-1,\,-1,\,-1,\,-1,\,-1,\,-1\}$, with
the sum of $\lambda$ and $\lambda^{-1}$, needed only in the
amplitude calculations, given as \be\label{lambda}
\la+\la^{-1}=-2\frac{1-f^2}{1+f^2}.\ee

We have then the amplitude in the NS-NS sector as \bea\label{AmpNSM}
\Gamma_{\rm NS-NS} &=&  _{\rm NS-NS}\bra{B^0}D\ket{B^8}_{\rm
NS-NS}\nn  &= &\frac{ n_0\,n_8\,V_1}{2 (8\pi^2\a')^{\frac{1}{2}}}
\int_0^\infty dt
e^{-\frac{Y^2}{2\pi\a't}}\,t^{-\frac{1}{2}}\left[\frac{\theta_3\left(\nu|i
t\right) \theta_3^3\left(\frac{1}{2}|i t\right)}{\theta_1\left(\nu|i
t\right)\theta_1^3\left(\frac{1}{2}|i t\right)}-
\frac{\theta_4\left(\nu|i t\right) \theta_4^3\left(\frac{1}{2}|i
t\right)}{\theta_1\left(\nu|i t\right)\theta_1^3\left(\frac{1}{2}|i
t\right)}\right], \eea where the parameter $\nu$ is defined via
$\la\equiv {\rm exp}(2\pi\,\ii\,\nu)$ with $\nu\in (0, 1/2]$ and
from (\ref{lambda}) we have $\sin \pi\nu = 1/\sqrt{1+f^2}$. The
amplitude in the R-R sector can also be calculated  as
\bea\label{AmpRM} \Gamma_{\rm R-R} = \,_{\rm R-R}\bra
{B^0}D\ket{B^8}_{\rm R-R} =
\frac{1}{2}n_0\,n_8\,V_1\,(8\pi^2\a')^{-\frac{1}{2}} \int_0^\infty
dt\,e^{-\frac{Y^2}{2\pi\a't}}t^{-\frac{1}{2}}, \eea where we have
employed the regularization for the fermionic zero modes given in
\cite{Billo:1998np} for the zero-mode matrix element, $ ^{(0)}_{\rm
R}\langle B_\psi^0,\eta^0 | B_\psi^8,\eta^8\rangle_{\rm R}^{(0)} =
-\frac{\d_{\eta^0\eta^8,-}}{\sqrt{1+f^2}} {\rm
Tr}\left(\Gamma^{0}\Gamma^{9}\right)$, which meets the
regularization requirement. Note that this R-R amplitude is indeed
the same as that in the absence of the flux, as anticipated, again
purely due to the massless modes. Therefore, the improved
interpretations discussed in section 2 for the attractive potential
in this sector continue to hold.

So the total amplitude $\Gamma = \Gamma_{\rm NS-NS}+\Gamma_{\rm
R-R}$ is now  \bea\label{AmpTM} \Gamma  &=& \frac{ n_0\,n_8\,V_1}{2
(8\pi^2\a')^{\frac{1}{2}}}\int_0^\infty dt
e^{-\frac{Y^2}{2\pi\a't}}\, t^{-\frac{1}{2}}
\left[1+\frac{\theta_3\left(\nu|i t\right)
\theta_3^3\left(\frac{1}{2}|i t\right)}{\theta_1\left(\nu|i
t\right)\theta_1^3\left(\frac{1}{2}|i t\right)}-
\frac{\theta_4\left(\nu|i t\right) \theta_4^3\left(\frac{1}{2}|i
t\right)}{\theta_1\left(\nu|i t\right)\theta_1^3\left(\frac{1}{2}|i
t\right)}\right]\nn &=& -
\frac{n_0\,n_8\,V_1}{(8\pi^2\a')^{\frac{1}{2}}}\int_0^\infty
dt\,e^{-\frac{Y^2}{2\pi\a't}}\,t^{-\frac{1}{2}}
\frac{\theta_1^4\left(\frac{1}{4} - \frac{\nu}{2}|\,\ii\,t\right)
}{\theta_1\left(\nu|\,\ii\,t\right)\theta_1^3\left(\half|\,\ii\,t\right)}\nn
&=& - \frac{n_0\,n_8\,V_1}{(8\pi^2\a')^{\frac{1}{2}}}\frac{\sin^4
\pi\left(\frac{1}{4} - \frac{\nu}{2}\right)}{\sin
\pi\nu}\int_0^\infty dt\,e^{-\frac{Y^2}{2\pi\a't}}\,t^{-\frac{1}{2}}
\nn & &~~~~~~~~~~~~~~~~~~\times
\prod_{n=1}^{\infty}\frac{\left(1-\bar{\la}
|z|^{2n}\right)^4\left(1-{\bar \la}^{-1}|z|^{2n}\right)^4}
{\left(1-\la |z|^{2n}\right)\left(1-\la^{-1} |z|^{2n}\right)\left(1+
|z|^{2n}\right)^6}, \eea where in the second equality we have made
use of the following fundamental Jacobian identity $
-2\theta_1^4\left(\left.\frac{\nu}{2}-\frac{1}{4}\right|\,\ii\,t\right)
=
\theta_1\left(\left.\nu\right|\,\ii\,t\right)\theta_1^3\left(\left.\half\right|\,\ii\,t\right)
+
\theta_3\left(\left.\nu\right|\,\ii\,t\right)\theta_3^3\left(\left.\half\right|\,\ii\,t\right)
-
\theta_4\left(\left.\nu\right|\,\ii\,t\right)\theta_4^3\left(\left.\half\right|\,\ii\,t\right)$
which can be obtained from the equation (iv) on page 468 in
\cite{Whittaker:1963cu} with certain choices of variables and also a
relation $\theta_1 (1 - \nu| it) = \theta_1 (\nu | it)$. In the
above $|z| = e^{- \pi t}$ and $\bar \la = {\rm exp}[2\pi i (1/4 -
\nu/2)]$. It is obvious that the amplitude vanishes only in the
absence of the flux, i.e., $\nu = 1/2$. Noticing that $\left(1-\la
|z|^{2n}\right)\left(1-\la^{-1} |z|^{2n}\right) = \left(1-2 \cos
2\pi\nu\, |z|^{2n}+|z|^{4n}\right)>0$, this amplitude is always less
than zero for the non-vanishing flux, therefore implying a repulsive
interaction as anticipated. This can further be understood as
follows: the interactions between the D0 and D6 and between the D0
and the D8 are both repulsive in the NS-NS sector while the
interaction from the R-R sector is purely from that between the D0
and D8. Also in the absence of the D6 in the bound state, the net
interaction vanishes. So now we must have a repulsive interaction
since the repulsive part is enlarged.

From the above calculations, it becomes clear that adding a flux,
electric or magnetic, to the D8 will not change the singular
behavior of the R-R potential which is due to massless modes. This
implies that if there is any change to the NS-NS potential, it must
be an overall factor and/or a change on the part due to massive
modes which will not be important for large t in the closed string
cylinder-diagram or small t in the open string annulus diagram.
Precisely in this region, the t-integration gives rise to the
singularity. This will remain so even for a more general flux or
fluxes. So we conclude that the non-threshold (F, D8) or (D6, D8)
bound state or a D8 carrying more general flux/fluxes will not exist
as an independent object just like the D8 carrying no flux.

\section{Summary}

In this paper, we address various subtle issues regarding the D0-D8
system. First we show that the potential in either the NS-NS or R-R
sector is actually divergent and can be expressed as a
brane-separation-independent divergent piece plus a
brane-separation-dependent finite piece. This divergent piece in
each sector is the interaction energy when the D0 and the D8 has
zero separation, therefore can be viewed as the zero-point energy in
this sector. The finite piece can be viewed as the Casimir energy
and in particular in the R-R sector it can be viewed as due to a
string created with its tension one half of that of a fundamental
string when the D0 moves away from the D8 at a distance.  With this,
we have improved the understanding of a fundamental string creation
as given in \cite{Kitao:1998vn} when the D0 crosses the D8 by
including the divergent piece in each sector before and after the
crossing. The other interpretation continues to hold if the
divergent piece is also included in the R-R potential.

We have also shown that the above improved interpretations regarding
the non-vanishing attractive R-R potential continue to hold even in
the presence of a flux, electric or magnetic. When the flux is
electric, we also resolve a subtle issue in section 3 on how to
implement properly the regularization of fermionic zero-modes given
in \cite{Billo:1998np} so that a meaning result can be obtained for
the R-R potential.

The persistence of divergence for the amplitude in each sector in
the presence of a flux, electric or magnetic, or even a more general
flux indicates that the non-threshold (F, D8) or (D6, D8) bound
state, or the D8 carrying a more general flux cannot exist as an
independent of object, either.

\vspace{.5cm}

\noindent {\bf Acknowledgements}

We would like to thank the anonymous referee for fruitful
suggestions which help us to improve the manuscript, to Zhao-Long
Wang and Shan-Shan Xu for useful discussion. We acknowledge support
by grants from the Chinese Academy of Sciences, a grant from 973
Program with grant No: 2007CB815401 and grants from the NSF of China
with Grant No:10588503 and 10535060.


\vspace{2pt}

\begin{thebibliography}{99}

\bibitem{polbooktwo}
  J.~Polchinski,
  ``Superstring Theory",
  Vol. 2, Page 175. Cambridge: Cambridge University Press (1998)

\bibitem{Polchinski:1995df}
  J.~Polchinski and E.~Witten,
  ``Evidence for Heterotic - Type I String Duality,''
  Nucl.\ Phys.\  B {\bf 460}, 525 (1996)
  [arXiv:hep-th/9510169].

\bibitem{Bergshoeff:2001pv}
  E.~Bergshoeff, R.~Kallosh, T.~Ortin, D.~Roest and A.~Van Proeyen,
  ``New Formulations of D=10 Supersymmetry and D8-O8 Domain Walls,''
  Class.\ Quant.\ Grav.\  {\bf 18}, 3359 (2001)
  [arXiv:hep-th/0103233].


\bibitem{Polchinski:1995prl}
  J.~Polchinski,
  ``Dirichlet-Branes and Ramond-Ramond Charges,''
  Phys.\ Rev.\ Lett.\ {\bf 75}, 4724 (1995)
  [arXiv:~hep-th/9510017].


\bibitem{Polchinski:1996fm}
  J.~Polchinski, S.~Chaudhuri and C.~V.~Johnson,
  ``Notes on D-Branes,''
  arXiv:hep-th/9602052;
  J.~Polchinski,
  ``Lectures on D-branes,''
  arXiv:hep-th/9611050.

\bibitem{Lifschytz:1996pl}
  G.~Lifschytz,
  ``Comparing d-branes to black-branes,''
  Phys.\ Lett.\ B {\bf 388}, 720 (1996)
  [arXiv:~hep-th/9604156].

\bibitem{Danielsson:1997wq}
  U.~Danielsson, G.~Ferretti and I.~R.~Klebanov,
  ``Creation of fundamental strings by crossing D-branes,''
  Phys.\ Rev.\ Lett.\  {\bf 79}, 1984 (1997)
  [arXiv:hep-th/9705084].

\bibitem{Bergman:1997gf}
  O.~Bergman, M.~R.~Gaberdiel and G.~Lifschytz,
  ``Branes, orientifolds and the creation of elementary strings,''
  Nucl.\ Phys.\  B {\bf 509}, 194 (1998)
  [arXiv:hep-th/9705130].


\bibitem{Billo:1998np}
  M.~Billo, P.~Di Vecchia, M.~Frau, A.~Lerda, I.~Pesando, R.~Russo and S.~Sciuto,
  ``Microscopic string analysis of the D0-D8 brane system and dual R-R
  states,''
  Nucl.\ Phys.\  B {\bf 526}, 199 (1998)
  [arXiv:~hep-th/9802088].

\bibitem{Polchinski:1995sm}
  J.~Polchinski and A.~Strominger,
  ``New Vacua for Type II String Theory,''
  Phys.\ Lett.\  B {\bf 388}, 736 (1996)
  [arXiv:hep-th/9510227].

\bibitem{Hanany:1996ie}
  A.~Hanany and E.~Witten,
  ``Type IIB superstrings, BPS monopoles, and three-dimensional gauge
  dynamics,''
  Nucl.\ Phys.\  B {\bf 492}, 152 (1997)
  [arXiv:hep-th/9611230].

\bibitem{Kitao:1998vn}
  T.~Kitao, N.~Ohta and J.~G.~Zhou,
  ``Fermionic zero mode and string creation between D4-branes at angles,''
  Phys.\ Lett.\  B {\bf 428}, 68 (1998)
  [arXiv:hep-th/9801135].


\bibitem{Lu:2009yx}
  J.~X.~Lu, B.~Ning, R.~Wei and S.-S.~Xu,
  ``Interaction between two non-threshold bound states,''
  Phys.\ Rev.\  D {\bf 79}, 126002 (2009)
  [arXiv:0902.1716 [hep-th]].


\bibitem{Friedan:1985ge}
  D.~Friedan, E.~J.~Martinec and S.~H.~Shenker,
  ``Conformal Invariance, Supersymmetry And String Theory,''
  Nucl.\ Phys.\  B {\bf 271}, 93 (1986).

\bibitem{Bianchi:1991eu}
  M.~Bianchi, G.~Pradisi and A.~Sagnotti,
  ``Toroidal compactification and symmetry breaking in open string theories,''
  Nucl.\ Phys.\  B {\bf 376}, 365 (1992).


\bibitem{Lu:2009se}
  J.~X.~Lu and S. -S. Xu,
  ``The open string pair-production rate enhancement by a magnetic flux,''
  JHEP {\bf 09}, 093 (2009) [arXiv:~0904.4112~[hep-th]].

\bibitem{Lu:2009th}
  J.~X.~Lu and S. -S. Xu,
  ``Remarks on D$_p$ $\&$ D$_{p - 2}$ with each carrying a flux,''
  Phys.\ Lett.\ B{\bf 680}, 387 (2009) [arXiv:~0906.0679~[hep-th]].

\bibitem{Witten:2000mf}
  E.~Witten,
  ``BPS bound states of D0-D6 and D0-D8 systems in a B-field,''
  JHEP {\bf 0204}, 012 (2002)
  [arXiv:~hep-th/0012054].

\bibitem{Mihailescu:2000dn}
  M.~Mihailescu, I.~Y.~Park and T.~A.~Tran,
  ``D-branes as solitons of an N = 1, D = 10 non-commutative gauge theory,''
  Phys.\ Rev.\  D {\bf 64}, 046006 (2001)
  [arXiv:hep-th/0011079].

\bibitem{Fujii:2001wp}
  A.~Fujii, Y.~Imaizumi and N.~Ohta,
  ``Supersymmetry, spectrum and fate of D0-Dp systems with B-field,''
  Nucl.\ Phys.\  B {\bf 615}, 61 (2001)
  [arXiv:hep-th/0105079].


\bibitem{Whittaker:1963cu}
  E.~T.~Whittaker and G.~N.~Watson,
  ``A Course of Modern Analysis", 4th Ed (reprinted).
  Cambridge: Cambridge University Press (1963)

\end{thebibliography}
\end{document}